\documentclass[12pt]{iopart}
\usepackage{iopams}
\usepackage{iopams,cite}
\usepackage{eurosym}
\usepackage{lipsum}
\usepackage{graphicx}
\usepackage{color}
\usepackage[normalem]{ulem}
\usepackage{lipsum}
\usepackage{tikz}
\usepackage{multicol}
\usepackage{float}
\usepackage{amssymb}
\usepackage{epsfig,color}

\begin{document}
\title{Bosonic impurity in a one-dimensional quantum droplet in the Bose-Bose mixture}
\author{F. Kh. Abdullaev$^{1,2}$ and R. M. Galimzyanov$^{1}$
\footnote{Corresponding author: ravil\_galimzyanov@yahoo.com}}
\address{$^1$ Physical-Technical Institute of the Academy of Sciences,
Chingiz Aitmatov street 2-b, 100084, Tashkent-84, Uzbekistan\\
$^2$ Physics Department, National University of Uzbekistan,
University street 4, 100174, Tashkent-174, Uzbekistan}
\begin{abstract}
 We study an impurity immersed in the mixture of Bose ultracold
gases in the regime where a quantum droplet exists. The quasi-one-
dimensional geometry is considered.  We find an effective
attractive potential which acts by the quantum droplet onto the
impurity. The bound states of the impurity in this potential are
investigated. These impurity bound states can provide potential
probes for the presence of quantum fluctuations effects on the
droplet properties. In the case of strong impurity-BEC coupling we
study the properties of the nonlinear local modes on the impurity
induced by the quantum fluctuations.
\end{abstract}
\date{\today}
\maketitle

\section{Introduction}
A lot of attention has recently been devoted to investigation of
self-supported quantum droplets (QD) in ultracold quantum
gases~\cite{Kadau,Schmitt,Ferrier,Chomaz,Cabrera,Semeghini}. They
represent new states of quantum liquids, and in the
distinction from droplets in the liquid helium, can exist at
densities that are lower by few orders. The possibility to tune
parameters of BEC in wide region open new directions in studying
of the QD.

The existence of these droplets in BEC with attractive two-body
interaction is connected with the stabilizing role being played by
quantum fluctuations ~\cite{Petrov}. The contribution to the ground
energy from quantum fluctuations was first calculated by Lee-Huang-Yang
~\cite{LHY} and has the repulsive character in 3D. In quasi-one- or
two-dimensional system it can be either attractive or repulsive,
depending on the density ~\cite{AstPetrov}. Theoretical investigation
of the dynamics of droplets in quasi-one-dimensional setup is carried
out in ~\cite{AM,Abd19,Abd19_1}. Balance between the residual mean
field attractive nonlinearity, produced by interaction of attractive
inter-species and repulsive intra-species, and repulsive LHY term leads
to existence of quantum droplets. At present time they are
theoretically predicted for Bose-Bose (BB) mixtures ~\cite{Petrov},
dipolar condensates ~\cite{Baillie}, Bose-Fermi (BF) mixtures
~\cite{FBmixture}, spin-orbit coupled BEC~\cite{SOC,Salashnich}. The
quantum droplets have recently been observed in BB
mixtures~\cite{Cabrera}, dipolar condensates~\cite{Ferrier,Schmitt},
and probably in BF mixtures~\cite{FBmixtureexp}. QD
formed by atoms of different atomic species have been reported in
\cite{DifSp}. Dynamics of the Faraday waves under action of the quantum
fluctuations is investigated in recent work ~\cite{Abd19}.

It should be noted that quantum droplets in ultracold gases
resemble ones in the  liquid helium. Interaction of an impurity
with the droplet in liquid helium leads to the rich phenomena as
creation clusters with confinement of single molecules inside or
on the surface of free droplets and so on~\cite{He4}.

Using this analogy we can expect that an impurity immersed into
the quantum droplet may have new interesting properties like
generation of quantum polarons (see
e.g.\cite{QP1,QP2,QP3,QP4,QP5,QP6}). Such an investigation has
recently been performed in the case of the fermionic impurity
embedded into the dipolar condensate~\cite{Wenzel1}.  It is shown
that this type of impurities provides unique probe for analysis of
the properties of the QD.

In the present work we will study another system, namely a neutral
Bose-impurity, immersed in a quasi-one-dimensional Bose-Bose
mixture. We study self-localization of the impurity and
find the bound state energies in the effective potential induced by a quantum droplet.
As well as we will investigate the case of strong coupling of an impurity with
BEC when localized state of the impurity is accompanied by strong deformation of the BEC ~\cite{Timmermans,Bruderer}.
Note that in these works the problem has been investigated without taking into account quantum
fluctuations.

The structure of the work is as follows: In section 2 we give basic equations
describing BB mixture and an impurity immersed there. In section 3 we find effective potentials and
calculate energy levels of bound states of an impurity at
different values of the system parameters. In section 4 the
case of strong coupling of the $\delta-$impurity with BEC is investigated. We
find nonlinear localized modes on the impurity and compare
analytical predictions with results of numerical simulations of
the full GP equation.

\section{Model}
The analysis is based on the model describing the interaction of
an impurity with the Bose-Einstein condensate. The system of
equations for wave functions of the one-component condensate and
an impurity,  has been earlier introduced in \cite{Gross,Timmermans}.

This system can be generalized for two component case, where components
correspond to different isotopes or hyperfine states. Dynamics of an impurity
immersed into the mixture of two Bose gases with taking into account
quantum fluctuations in the LHY form~\cite{AstPetrov} is described by
the coupled system of equations for wave functions of the BB mixture
$\Psi_1,\Psi_2$ and the impurity $\Phi $
\begin{eqnarray}\label{eq0}
&&i\hbar\Psi_{1,t}+\frac{\hbar^2}{2m_1}\Psi_{1,xx}- (g_{11}|\Psi_1|^2 + g_{12}|\Psi_2|^2)\Psi_1+ \nonumber\\
&&\frac{\sqrt{m}g_{11}}{\pi\hbar}(g_{11}|\Psi_1|^2 + g_{22}|\Psi_2|^2)^{1/2}\Psi_1 -g_{IB1}|\Phi|^2\Psi_1=0,\\
&&i\hbar\Psi_{2,t}+\frac{\hbar^2}{2m_2}\Psi_{2,xx}- (g_{22}|\Psi_2|^2 + g_{21}|\Psi_1|^2)\Psi_2 +\nonumber\\
&&\frac{\sqrt{m}g_{22}}{\pi\hbar}(g_{11}|\Psi_1|^2 + g_{22}|\Psi_2|^2)^{1/2}\Psi_2 -g_{IB2}|\Phi|^2\Psi_2=0,\\
&&i\hbar\Phi_{t}+\frac{\hbar^2}{2m_I}\Phi_{xx}-(g_{IB1}|\Psi_1|^2+g_{IB2}|\Psi_2|^2)\Phi=0,
\end{eqnarray}
where notation $_{,y}$ is used for the partial derivative with
respect to a generic variable $y$,
$g_{ij}=2\hbar\omega_{\perp}a_{ij}$, with $\omega_{\perp}$ being
the transverse frequency of the trap (BECs and impurities are
supposed to have the same transverse frequency) $a_{ij}$ is the
intra- and inter-species atomic scattering lengths, $a_{IB1,2}$ is
the impurity-boson scattering lengths. The scattering lengths can be
tuned by using the Feshbach resonance technique~\cite{FR}. Hereafter we
assume $g_{11}=g_{22}=g$, $m_1 = m_2 = m_I = m$ and $g=|g_{12}|+\delta g$,
$g_{IB1}=g_{IB2}=g_{IB}$. Then we can suppose $\psi_1 = \psi_2 =\psi$
so the system of equations takes the form
\begin{eqnarray}\label{eq1}
i\hbar\Psi_t + \frac{\hbar^2}{2m}\Psi_{xx}- \delta g|\Psi|^2\Psi - g_{IB}|\Phi|^2\Psi  + 
g_{QF}|\Psi|\Psi=0, \\
i\hbar \Phi_{t} + \frac{\hbar^2}{2m}\Phi_{xx} -2g_{IB}|\Psi|^2\Phi
= 0,
\end{eqnarray}
where $g_{QF}=\sqrt{2m}g^{3/2}/(\pi \hbar)$. Note that in the case of
one dimension, quantum fluctuations lead to the effectively
attraction effect. Then, for existence of the quantum droplet in
this limit, we should choose the small residual two-body repulsion i.e.
$\delta g > 0$.  Defining characteristic units of length $x_0$, time $t_0$, wave function $\Psi_0$ and energy $E_0$ (see e.g.~\cite{AM})
\begin{equation}\label{dim}
x_0 = \frac{\pi \hbar^2\delta g_0^{1/2}}{\sqrt{2}mg^{3/2}}, \ t_0=\frac{\pi^2\hbar^3\delta g_0}{2mg^3}, \  \Psi_0=\frac{\sqrt{m}g^{3/2}}{\pi\hbar \delta g_0},\ E_0=\frac{2mg^3}{\pi^2\hbar^2\delta g_0},
\end{equation}
where $\delta g_0$ is a fixed value of the detuning of inter- and
intra- species scattering lengths, and rescaling $x'=x/x_0, \
t'=t/t_0, \  \psi'= \psi/\psi_0$,  we get the following
dimensionless system of equations (the primes are omitted)
\begin{eqnarray}
i\psi_t + \frac{1}{2}\psi_{xx} -\gamma |\psi|^2\psi + |\psi|\psi + f_1|\phi|^2\psi =0, \label{eq3eq4:one}
\\
i\phi_t + \frac{1}{2}\phi_{xx} + f_2|\psi|^2 \phi =0,
\label{eq3eq4:two}
\end{eqnarray}
where $\gamma = \delta g/\delta g_0$,
$f_1=-N_{IB}g_{IB}/(N_b \delta g_0), \ f_2=-2 g_{IB}/\delta g_0$.

Let us consider the case when $f_1=f_2=0$ that corresponds to  a
uncoupled system. Exact solution for the droplet ($\gamma =1$) has
the form
\begin{equation}\label{droplet}
\psi(x)=\frac{3\mu e^{i\mu t}}{1+\sqrt{1-9\mu/2 }\cosh(\sqrt{2\mu}\,x)}.
\end{equation}
The number of atoms in this wavepacket
\begin{equation}\label{Nmu}
N=\int_{-\infty}^{\infty} |\psi(x)|^2 dx = \frac{4}{3}\left[ \ln\left(\frac{\sqrt{9\mu/2
}+1}{\sqrt{1-9\mu/2}}\right)-\sqrt{\frac{9\mu}{2}}\right].
\end{equation}
For small $N\ll1$ the droplet profile is well described by the
Gaussian function, and for $N\gg1$ the profile turns into the flat top one,
extending with increasing  $N$. When $N\ll 1$, chemical potential $\mu \approx 0.382N^{2/3}$
and $\mu \approx 2/9$ for $N\gg 1$~\cite{AM}.

\section{Effective potential and energy levels for an impurity trapped by the quantum droplet}
Now we consider the case when the back action of the impurity on the
BEC can be neglected (i.e. $f_1=0$). From the equation
Eq.~(\ref{eq3eq4:two}) one can see that the impurity is in
the effective potential induced by a quantum droplet
\begin{equation}\label{Veff}
V_{eff}(x)=-f_2|\psi(x)|^2=
-f_2\frac{9\mu^2}{(1+\sqrt{1-9\mu/2}\cosh^2(\sqrt{2\mu}\,x))^2} .
\end{equation}
There are  two limits

(i) Small $N$, when the profile of the droplet wave function $\psi$ is narrow.

(ii) Large $N$, when the potential is described by the flat-top
function and so can be approximated by a rectangular potential well.
It should be noted that the approach using an effective potential for the impurity
systems in BEC has recently been used in works\cite{EP1,EP4,EP5}.

\subsection{Effective potential and energy levels when $\gamma = 0 \ (i.e. \  \delta g = 0)$}
In this important case the residual mean field term vanishes,
i.e. the inter- and intra-species interactions
balance each other.
The leading effect in the formation of the droplet is given
by the LHY term, and effects of quantum fluctuations dominate.
Exact solution for a quantum droplet has the form
~\cite{Abd19_1,Salashnich}
\begin{equation}\label{phi0}
\tilde{\psi}=\frac{3\tilde{\mu} e^{i\tilde{\mu} t}}{2\cosh^2 (\sqrt{\tilde{\mu} /2}\,x)},
\end{equation}
where the chemical potential $\tilde{\mu}$ may take values $0<
\tilde{\mu} <\infty$.

Interaction between the quantum droplet (\ref{phi0}) and an impurity is determined by an effective potential
\begin{equation}\label{V0eff}
V_{0,eff}(x)=
-f_2\frac{9\tilde{\mu}^2}{4\cosh^4(\sqrt{\tilde{\mu} /2}\,x)} .
\end{equation}
The energy levels for this potential is difficult to find analytically,
therefore we will use an approximation of it by a potential of the solvable form.

\begin{figure}[htb]
\centering
\includegraphics[width=6.0cm,height=8cm,angle=-90,clip]{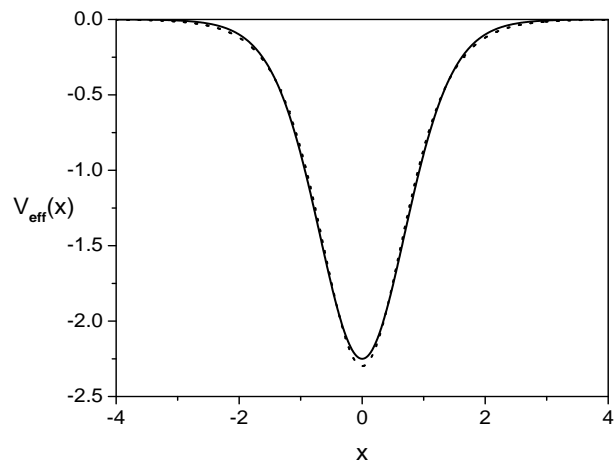}\quad
\caption{Effective potential for the droplet Eq.~(\ref{phi0}) and
its approximated variant $V_{0,appr}(x)$ allowing to find analytically all
eigenvalues and eigenfunctions. Fitting parameters $a = 1.022$, $b = 1.523$.} \label{V0ef}
\end{figure}

It turns out that the function $y(z) = 1/\cosh^4(x)$ is well
approximated by the expression $a/\cosh^2(bx)$ with fitting parameters $a$ and $b$.
Then effective potential $V_{0,eff}(x)$ can be well approximated by new one
\begin{equation}\label{approx}
V_{0,eff}(x) \approx V_{0,appr}(x)=
-af_2\frac{9\tilde{\mu}^2}{4\cosh^2(\sqrt{\tilde{\mu} /2}(bx))}.
\end{equation}
Effective potential $V_{0,eff}(x)$ and its
approximated variant $V_{0,appr}(x)$ are shown in Fig.~\ref{V0ef}.
Here dot line is for exact effective potential $V_{0,eff}$ Eq.~(\ref{V0eff}) and
solid line is for its fitted variant. One can see that the curves are in a good agreement with each other.

Introducing new parameter $$\lambda = \frac{1}{2} + \sqrt{\frac{1}{4} + f_2 \frac{9 \tilde{\mu}a}{b^2}},$$ approximating potential $V_{0,appr}(x)$ can be transformed to the modified
P\"oschl-Teller potential ~\cite{Flugge}, which takes the following form
\begin{equation}\label{PTeller}
V_{0,Poschl-Teller} = - \frac{1}{2} \frac{(\sqrt{\tilde{\mu} /2}b)^2 \lambda (\lambda - 1)}{\cosh^2(\sqrt{\tilde{\mu} /2}\,b x)}.
\end{equation}
Everywhere fitting parameters are: $a = 1.022$ and $b = 1.523$.

All this makes it possible to get approximated analytical
expressions for $n$ levels in the potential $V_{0,eff}$
Eq.~(\ref{V0eff}),
\begin{eqnarray}\label{levels}
E_n = - \frac{1}{2} (\sqrt{\tilde{\mu} /2}\,b)^2 (\lambda - 1 - n)^2 ,
\end{eqnarray}
where $n < \lambda - 1$ ~\cite{Flugge}.


The distance between energy levels can be found from Eq.~(\ref{levels}) as
\begin{equation}
\Delta E_n = E_n - E_{n-1} = b^2\frac{\tilde{\mu}}{2}\left(\lambda -n -\frac{1}{2}\right), \ n< \lambda -1.
\end{equation}

One can conclude that properties of the bound states are strongly
defined by quantum fluctuations effects, so measuring the energy
spectrum we can probe the quantum fluctuations in BEC.

\subsection{Effective potential and energy levels when $\gamma \neq 0 \ (i.e. \  \delta g \neq 0), \ f_1 = 0$ and $f_2 = 10$}
Let us consider the case when
$\delta g \neq 0$. Here it is difficult to find analytical
expression for all energy levels, so we will perform numerical
simulations.
\begin{figure}[htb]
\centering
\includegraphics[width=6.0cm,height=8cm,angle=-90,clip]{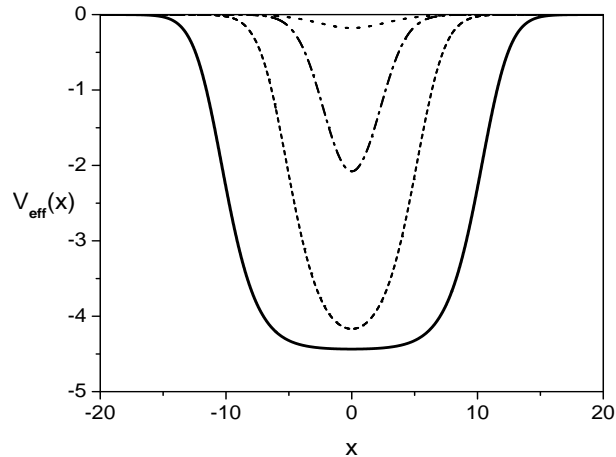}\quad
\caption{Forms of the droplet effective potential Eq.~(\ref{Veff})
at different chemical potentials $\mu$. Dot line is for $\mu =
0.08$, dash dot line is for $\mu = 0.2$, dash line is for $\mu =
0.222$ and solid line is for $\mu = 0.222222$. Everywhere $f_2 =
10$.} \label{Vef}
\end{figure}
In our calculations all physical values are dimensionless ones in
correspondence with Eqs. (\ref{dim}). In Fig. \ref{Vef} the forms
of the effective potential induced by
the quantum droplet consisting of $N$ atoms are shown ($N$ is directly determined by the parameter $\mu$). One
can see that the closer $\mu$ to the threshold value $\mu = 2/9$
the closer the form of the droplet effective potential to a wide
rectangular well.
\begin{figure}[htb]
\centering
\includegraphics[width=6.0cm,height=8cm,angle=-90,clip]{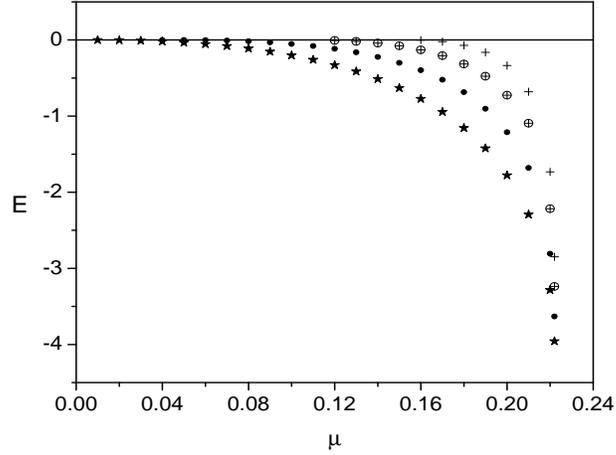}\quad
\caption{The impurity energy levels at different $\mu$: asterisks
are for ground states, filled circles are for the second energy
levels, circles with crosses are for the third energy levels and
crosses are for the fourth energy levels. Everywhere $f_2 = 10$.}
\label{Eigval}
\end{figure}
Fig. \ref{Eigval} depicts energy levels of the impurity depending
on different values of the chemical potential $\mu$.
\begin{figure}[htb]
\centering
\includegraphics[width=6.0cm,height=8cm,angle=-90,clip]{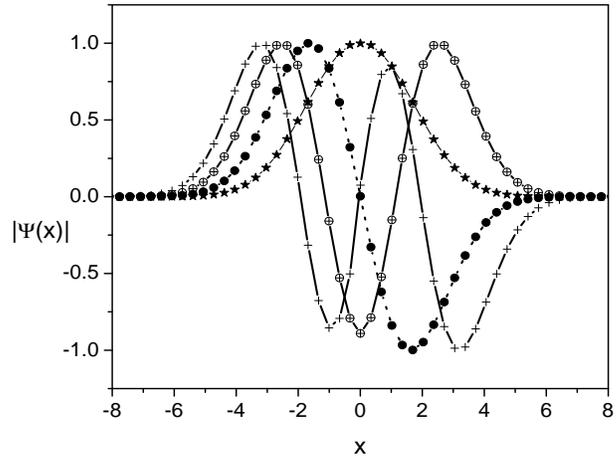}\quad
\caption{Stationary wave packets of the impurity for $\mu = 0.222$ corresponding
to $N = 4.2$. The line with asterisks is for the ground state wave
packet, the line with filled circles is for a wave packet of the
second energy level, the line with circles  containing crosses
inside is for the wave packet of the third energy level and the
line with crosses is for the wave packet of the fourth energy
level. The wave packets are of the same amplitude and they are not
normalized.} \label{Eigvect}
\end{figure}
 The first four eigenfunction are shown in Fig. \ref{Eigvect}
for the case $\mu = 0.222$ (dimensionless number of atoms, $N =
4.2$). The eigenfunctions are not normalized and have the same
amplitude.




Now we turn to the case when number of atoms $N$ in the system is
large. The dependence of $N$ on the value of chemical potential
$\mu$ is determined by Eq.~(\ref{Nmu}). At large number of atoms
$\mu \rightarrow 2/9$ and may be expressed in terms of $N$
as in ref.~\cite{AM}
$$\mu \approx \frac{2}{9}\left[ 1-4\exp(-2-\frac{3}{2}N)\right].$$
As shown in work~\cite{Abd19_1}, stationary quantum droplet
solution Eq.~(\ref{droplet}) may be presented as sum of two kink
solutions
\begin{equation}\label{TransFio}
\psi(x)=
\sqrt{\frac{\mu}{2}}\left[\tanh\left(\sqrt{\frac{\mu}{2}}\left(x+x_0(\mu)\right)\right)+\tanh\left(\sqrt{\frac{\mu}{2}}\left(-x+x_0(\mu)\right)\right)\right]
e^{i\mu t},
\end{equation}
where $$x_0(\mu)=\frac{1}{\sqrt{2\mu}}\mbox{arctanh}\left(\sqrt{\frac{9\mu}{2}}\right).$$

Effective potential induced by the quantum droplet is determined
by Eq.~(\ref{Veff}), $V_{eff}(x)= -f_2|\psi(x)|^2$. When number of
atoms is large and $N\gg1$ with $\mu\rightarrow 2/9$, the form of
the potential $V_{eff}(x)$ turns into a rectangular well potential
with the fixed depth $V_0 = -f_2 2\mu$ and the width $2x_0=2.25N+3$.
It should be reminded that $N$ and $x_0$ are dimensionless values in
accordance with Eq.~(\ref{dim}).

Transitions between first nearest deep energy levels in the
potential are determined as
\begin{equation}\label{transitions}
\Delta E_n = E_n - E_{n-1} = f_2\frac{\pi}{8x_0^2}(2n+1),
\end{equation}
where $n$ is the level number.

The level spectrum may be measured by periodical variation in time
of the scattering length $a_{IB}=a_0 + a_1\sin(\omega t)$ with the
frequency $\omega =\Delta E/\hbar$, which induces resonant
transitions between levels ~\cite{Wenzel1,Gross,Abd2000} with
$\Delta E \approx (50 \div 100)Hz$ (see below). Also for
measurements of the impurity spectrum, the radio-spectroscopy
method ~\cite{Shash} can be used.

Let us estimate the parameters for the experiment. We can consider
binary $^{85}$Rb atoms BEC in the cigar type quasi-one-dimensional trap
with the transverse oscillator length $l_{\perp} \approx 0.6\ \mu m$.
The atomic scattering lengths can be chosen as $a_{11}=a_{22}=2000a_0,
\ a_{12} \approx -(0.95 \div 0.99)a_{11}$, that corresponds to $f_2 =20
\div 100$. In this case the characteristic scale of the length is $l_s
\sim (1 \div 0.4)\ \mu$ m, the time scale is $t_s \sim (0.8 \div 0.2)\
$ ms and the atoms number is $N_s \approx 40-450$~\cite{Abd19_1}. The
energy scale $E_0=\bar h/t_0$ is of order $(50 \div 100)Hz$ for the
above data.

\section{Strong coupling of an impurity with BEC
}
Next we consider effect of strong coupling on the BEC deformation
by an impurity. In the mean field approximation for one-component
BEC, self-trapping properties of impurities for strong attractive
and repulsive cases have been investigated in ~\cite{Bruderer}. We
can approximate the impurity action by considering it as
$|\phi(x)|^2 =\delta(x)$, i.e. as the  $\delta$-impurity. This
approximation is valid when the localization scale $l \ll l_h$,
where $l_h$ is the healing length $l_h$.
The $\delta$-impurity is placed at the point $x=0$. So the BEC
wave function is described by a stationary GP equation, which
follows from Eq.~(7) (note that $\Psi(x,t)=\psi(x)\exp(i\mu t)$)
\begin{equation}\label{deltaEq}
-\mu \psi + \frac{1}{2}\psi_{xx}-|\psi|^2u +|\psi|\psi -
A\delta(x)\psi=0,
\end{equation}
where $A$ is the strength of the $\delta$-potential.
We look for a solution in the form
\begin{eqnarray*}\label{wfdelta}
\psi(x) = \psi_0(x+x_0), \ x<0,   \\
\psi(x) = \psi_0(x-x_0), \  x>0,
\end{eqnarray*}
where $\psi_0(x)$ is a stationary solution of
Eq.~(\ref{eq3eq4:one})
\begin{equation}\label{sol}
\psi_0(x)=\frac{3\mu}{1+\sqrt{1-9\mu/2}\cosh(\sqrt{2\mu}\,x)}.
\end{equation}
for some given value of the chemical potential $\mu$.

Integrating Eq.~(\ref{deltaEq}) one time around the point $x=0$ we obtain
\begin{equation}\label{DelCond}
\psi_x(+0) - \psi_x(-0)= -2A\psi(x_0).
\end{equation}
Substituting  solution ~(\ref{sol}) into Eq.~(\ref{DelCond}) , we
come
to an equation for the parameter $x_0$
\begin{equation}\label{Eqx0}
\frac{\sqrt{2\mu}\sqrt{1-9\mu/2}\sinh(\sqrt{2\mu}\,x_0)}
{1+\sqrt{1-9\mu/2}\cosh(\sqrt{2\mu}\,x_0)} = A.
\end{equation}
This equation has exact solution for a localized mode with respect to $x_0$
\begin{equation}\label{parx0}
x_0=\sqrt{2/\mu}\hspace{0.1in}\mbox{arctanh}\left(\sqrt{2/\mu}\sqrt{1-9\mu/2}\hspace{0.1in}\frac{-1+\sqrt{1+\frac{A^2}{4/9-2\mu}}}{A(1-\sqrt{1-9\mu/2})}\right).
\end{equation}

From Eq.~(\ref{Eqx0}) follows the inequality $$|A| < \sqrt{2\mu}$$
which determines an existence domain of stationary solutions of
Eq.~(\ref{deltaEq}).

The number of atoms in the nonlinear localized mode is determined as
\begin{eqnarray}\label{Ncmu}
N_c=\int_{-\infty}^{\infty} |\psi(x)|^2 dx = \nonumber \\ 2\sqrt{2\mu}\left[2\sqrt{\frac{2}{9\mu}}\left(\mbox{arctanh}\left(\sqrt{\frac{a}{a+b}}\right)+\mbox{arctanh}\left(\sqrt{\frac{a}{a+b}}\mbox{tanh}(x'_0)\right)\right)\right.\nonumber \\
\left.-\left(1+\frac{b\sinh(x'_0)\cosh(x'_0)}{a+b\cosh^2(x'_0)}\right)\right],
\end{eqnarray}
where $$a= 1 - \sqrt{1-9\mu/2},  \ \  b= 2\sqrt{1-9\mu/2}, \ \
x'_0= \sqrt{\mu/2}x_0.$$

The dependence $N_c$ versus $\mu$ is depicted in Fig.~\ref{Ncmu}
for two values of the strength, $A=+0.15$ (dot line) and
$A=-0.15$. According to the Vakhitov-Kolokolov (VK)
criterion~\cite{VK}, a positive slope of the curve $N_c(\mu)$
corresponds to the stability  of the stationary solution
$\psi(x)$. One can see that the slope is positive over the entire
range of values of the parameter $\mu$ for both cases. Thus it
should follow that both our solutions must be stable. Further we
will demonstrate that the problem of stability is more
complicated.

\begin{figure}[htb]
\centering
\includegraphics[width=6.0cm,height=8cm,angle=-90,clip]{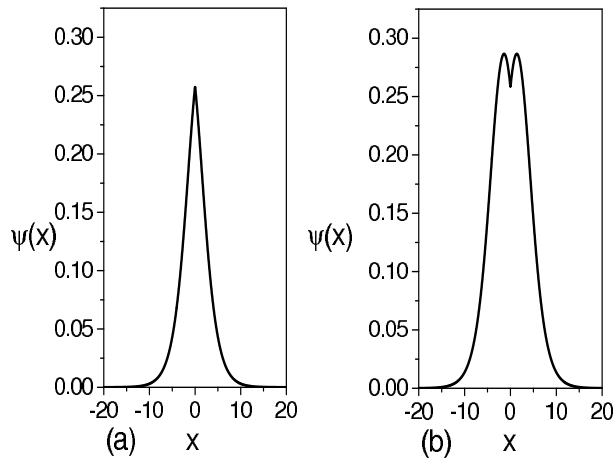}\quad
\caption{Stationary solutions of Eq.~(\ref{deltaEq}): (a)
double-peaked solution for repulsive delta-potential with
$A=0.15$; (b) single peaked solution for attractive
delta-potential with $A=-0.15$. The chemical potential $\mu=0.15$
in both cases.} \label{delSol}
\end{figure}

\begin{figure}[htb]
\centering
\includegraphics[width=7.0cm,angle=0,clip]{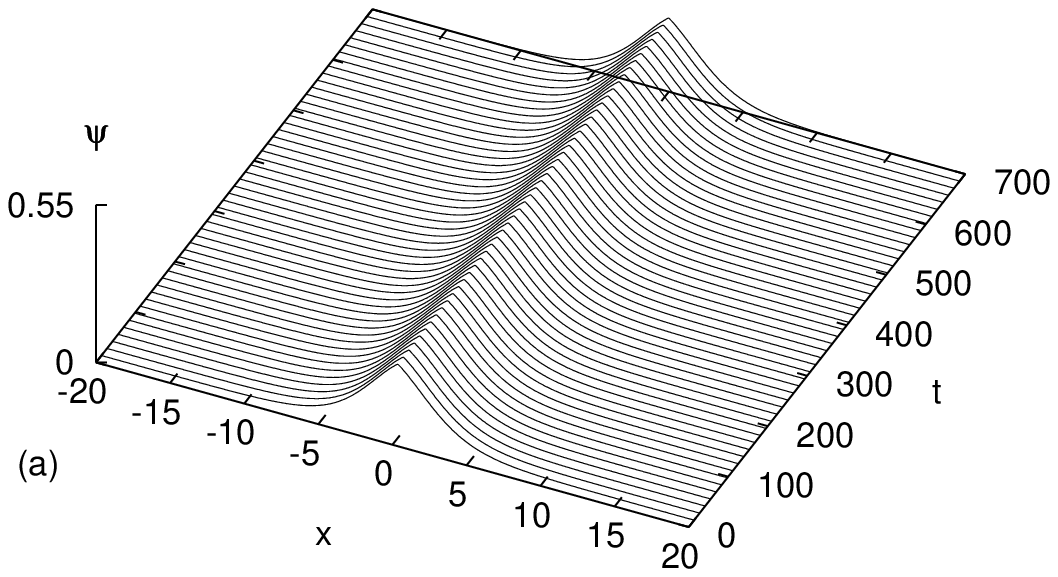}\quad
\includegraphics[width=7.0cm,angle=0,clip]{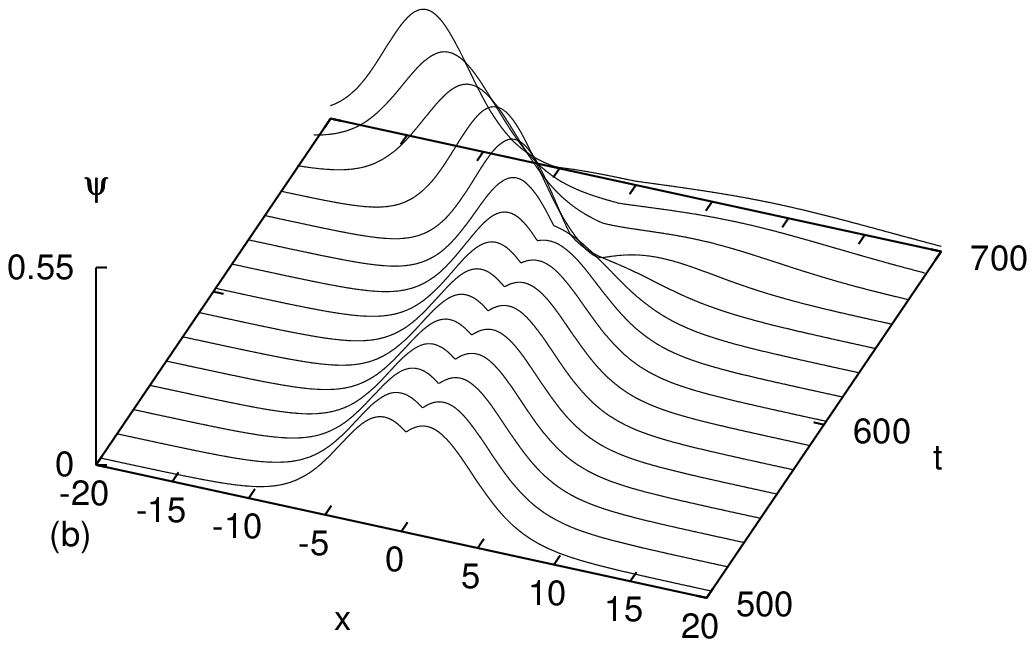}\quad
\includegraphics[width=7.0cm,angle=0,clip]{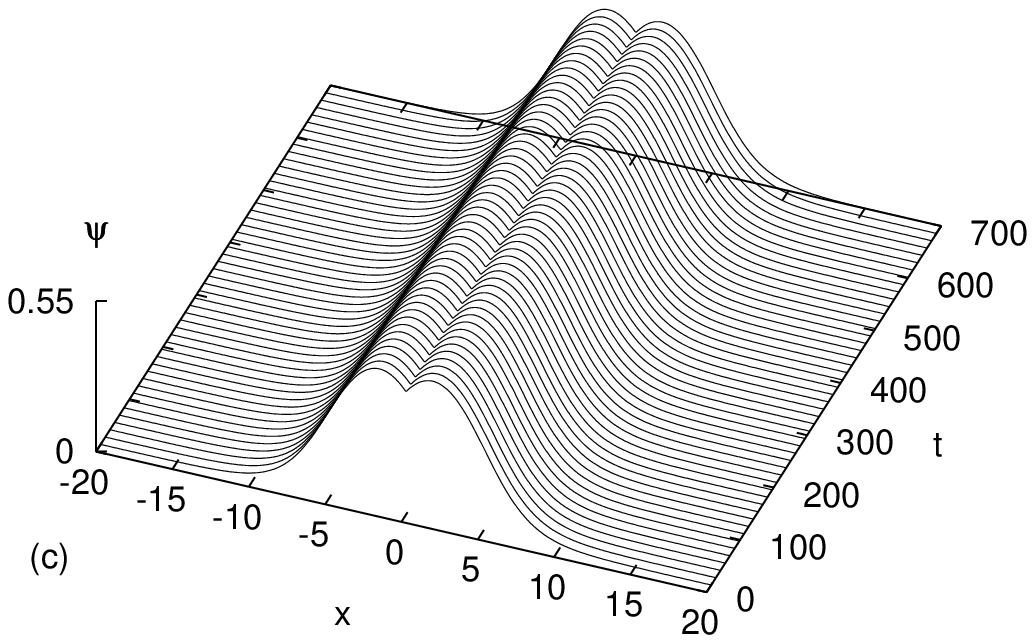}\quad
\caption{Stationary solutions of Eq.~(\ref{deltaEq}). Evolution of
stationary solutions: (a) stable solution for the attractive
$\delta$-potential with $A=-0.15$ and $\mu=0.15$; (b) unstable
solution for the repulsive delta-potential with $A=0.15$ and
$\mu=0.15$; (c) stable solution for the same repulsive
delta-potential with $A=0.15$ but another $\mu=0.21$.}
\label{EvolSol}
\end{figure}

In Fig. ~\ref{delSol} stationary solutions of the governing Eq.~(\ref{deltaEq}) are shown: (a) a single peaked solution for the attractive impurity, $A=-0.15$ and (b) a double-peaked one for the repulsive impurity, $A=+0.15$.
In both cases the chemical potential $\mu=0.15$.
The region of parameters exists (e.g. $A<0, \ \  A^2/2<\mu<2/9$),
where the solution is stable. Single-peaked solutions are
localized on an impurity  corresponding to the attractive
$\delta$-potential. Such solution with $A=-0.15$ ($\mu=0.15$) is
depicted in Fig.~\ref{EvolSol}a. It remains stable in entire
time-interval of the numerical simulations.


Double-peaked ones are localized on an impurity corresponding to
the repulsive $\delta$-potential. The numerical simulations show
that the stability of these solutions, as was mentioned above, is
complicated. For instance, as shown in Fig.~\ref{EvolSol}c, the
solution for positive $A=+0.15$  is stable at the value of the
chemical potential $\mu=0.21$.

The Vakhitov-Kolokolov slope criterion gives necessary stability
condition for the localized solution. But this condition is
insufficient for the case when the solution is centered at the
potential maximum. The wavepacket can drift away from its initial
location and the drift instability will develop~\cite{Fibich}.
We have performed numerical simulation for the case when $A=+0.15,
\ \ \mu=0.15$ and observed the drift instability of the solution
for a repulsive impurity (see Fig.~\ref{EvolSol}b).
 Detailed analysis of this instability requires separate investigation of the spectral condition~\cite{Gril} for Eq.~(\ref{deltaEq}).
\begin{figure}[htb]
\centering
\includegraphics[width=6.0cm,angle=-90,clip]{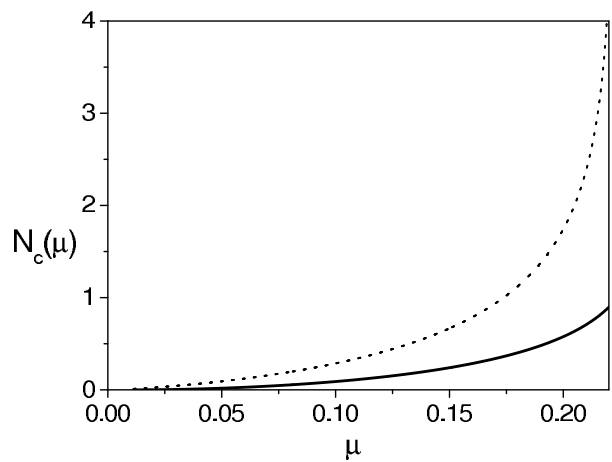}\quad
\caption{The dependence $N_c$ versus $\mu$: solid line is for the potential strength $A=-0.15$; dot line is for
the $A=+0.15$.} \label{Ncmu}
\end{figure}



%
.
\section{Conclusion}
In conclusion we have studied properties of a neutral bosonic
impurity immersed into a quantum droplet. We have found an
effective potential acting on the impurity from the side of the
quantum droplet. We have calculated energy levels for bound states
of the impurity in this potential for cases of dominating quantum
fluctuations as well as effects of the mixed mean field and
quantum fluctuations. In the case of strong coupling of the
impurity with BEC   we have shown the existence of the nonlinear
local mode of the impurity produced by the balanced mean field
nonlinearities and quantum fluctuations. We have found the
strength threshold of the impurity for existence of  nonlinear
localized modes in BEC. In our numerical simulations we have
demonstrated appearance of drift instability of the solution for
the repulsive impurity.

In future, it will be interesting to study the bound states of an impurity immersed in a three-dimensional droplet as well as
interaction of an impurity with quantum two-dimensional vortices. The problem of stability and the existence of nonlinear
localized modes of 2D and 3D systems is also required consideration.

\section*{Acknowledgments}
\label{sec:ack}

The authors acknowledge support by Fund for Fundamental
Researches of the Uzbekistan Academy of Sciences (Award No FA-F2-004).
\\

\end{document}